# Fractal analysis of $BaF_2$ thin films deposited on different substrates

Kavyashree[1], R. K. Pandey[2], R. P. Yadav[1*], Manvendra Kumar[2], A. K. Mittal[4],

A. C. Pandey[2], S. N. Pandey[1]

[1]Department of Physics, Motilal Nehru National Institute of Technology, Allahabad – 211004, India

[2]Nanotechnology Application Centre, University of Allahabad, Allahabad, 211002, India

[3]Department of Physics, University of Allahabad, Allahabad 211002, India

## Abstract

Barrium fluoride ($BaF_2$) thin films were prepared by electron beam evaporation technique at room temperature, on glass, Silicon and Aluminum substrates having thickness of 20 nm each. Its structural property and surface morphologywere studied using glancing angle X-ray diffraction (GAXRD) and atomic force microscopy (AFM) respectively. It was found that grain size, average surface roughness and interface width changes with different substrates. Higuchi's algorithm is applied for the fractal measure on AFM images.It was observed found that the fractal dimension varied from substrate to substrate.

[*]Corresponding author: aurampratap@gmail.com

# 1. Introduction

Barium fluoride has a fluorite cubic lattice structure in which a cube of $Ba^{2+}$ ions is surrounded by a cube of $F^-$ ions.It is a large band gap material having high electrical resistance so it behavesas aninsulating material. It is used as the optimal material for the transmission window in the optical instruments. It has also been used in infrared spectroscopy as window, mostly in the field of fuel oil analysis [1].It is also useful in high-power $CO_2$ laser applicationsscintillatorsfor the detection of gamma rays, X-rays,high energy particles, etc. [2]. It has also applications as a molten bath component in aluminium raffination, as flux and preopacifying agent and as a component of welding rod coatings and welding powders.

Surface plays a very significant role in several physical developments of materials and in different fields of modern day science and technology.Various properties such as optical, mechanical, magnetic, electrical, tribological, etc.depend on the microstructureand surface morphologies of thin films [1-3]. Precise control on the growth of the surface morphology and microstructure can be important aspects to produce the films for desired applications.As the surface morphology affectsmany properties, so beforehand investigation of surfaces are highly beneficial for understanding the surface for the usage of its further applications. In the current scenario, fractal analysis for the characterization of surface morphology is highly important to know the complexity of the surface[4-9]. The fractal analysis can be utilized to extract various types of information in contrast to the traditionalanalysis [10-16]. So, fractal analysis is broadly applied for the description of complex surfaces and improving our knowledge how the surface geometry affects the physical properties of the system[1, 3, 5, 17].

## 2. Experimental details and fractal assesment

BaF$_2$ thin films were deposited on the glass, Silicon (Si) and Aluminium (Al) substratesof 20 nm thickness each electron beam evaporation techniques at room temperature and in a vacuum environment of ~ 10$^{-6}$ mbar. The rate of deposition was 0.4-0.5nm/sec. Prior to the deposition, all the substrates were thoroughly cleaned. While performing the experiment, the thicknesses of the films were continuously monitored by quartz crystal monitor mounted in deposition chamber for the purpose. After deposition,all the samples were subjected to glancing angle X- ray diffraction (GAXRD) measurements to investigate their crystalline and structural properties. For this purpose Bruker AXS D8 advanced diffractometer being operated at 40 kV voltage and 40 mA current was used. The GAXRD patterns were recorded in the 2θ range of 20°- 60°. The surface morphology of the thin films were characterized under ambient condition using atomic force microscopy (AFM) images of size 2.0 µm x 2.0 µm digitized into 512 pixels x 512 pixels.

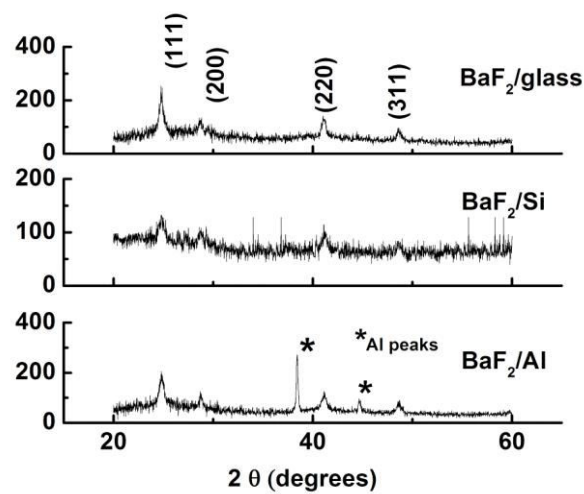

Figure 1: XRD pattern of BaF$_2$ thin film deposited on the glass, Si and Al substrates

In thin films, determination of the roughness is an important property for describing the surface morphology. Experimentally, the surface height is measured over a discrete lattice. Average roughness ($R_a$), and interface widths (w) of the surface are important parameters in quantitatively analyzing the characteristics of the surface morphology of thin films. We computed the values of average roughness using the relation $R_a = \langle |z(i,j) - \langle z(i,j) \rangle| \rangle$ and interface width by $w = \sqrt{\langle [z(i,j) - \langle z(i,j) \rangle]^2 \rangle}$, where $\langle z(i,j) \rangle$ represents the mean value of the heights over a square surface of side L[6]. Here, $z(i,j)$ denotes the surface height at point $(i,j)$. $R_a$ and $w$ characterize the global roughness of a surface but does not characterize surface correlation, which is an important characteristic of the surface. Fractal analysis can provide richer characterization of the surface morphology and deeper insight into the surface creation processes

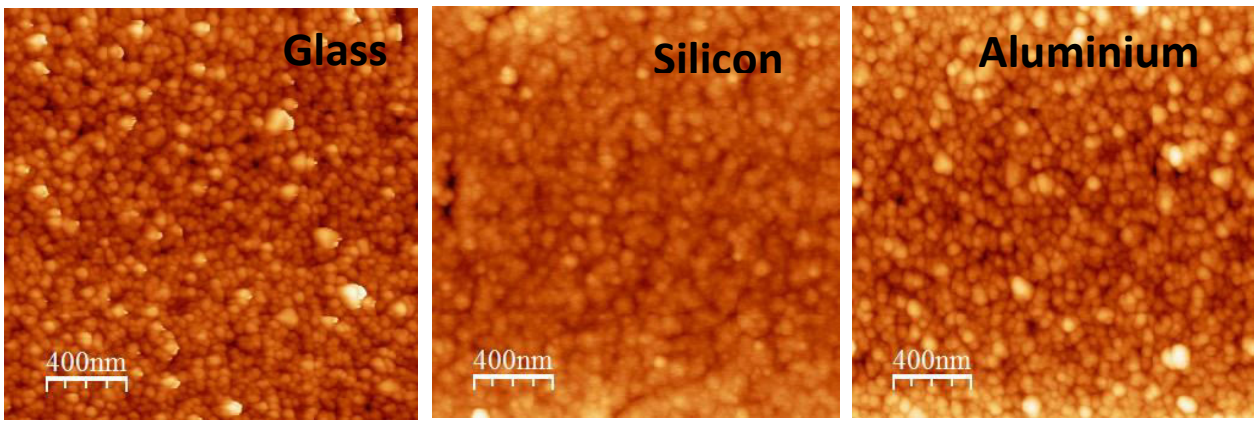

Figure 2: The AFM images of BaF$_2$ thin film surface deposited on the glass, Si and Al substrates.

Higuchi algorithm [7] is very efficient technique for computing the fractal dimension of a vertical as well as horizontal section of surfaces from the discretized height data recorded by AFM data. In this process, we construct $k$ subsequence of original discretized height data. Detailed about the algorithm is reported in Ref.[8]. The estimated average curve length ($L(k)$) using Higuchi algorithm follow the power law scaling $L(k) \sim k^{-D_f}$, where $D_f$ is the Higuchi fractal dimension [7]. The value of the fractal dimension $D_f$ is calculated by the least square linear best fit to a $\log L(k)$ versus $\log k$ curve.

## 3. Result and discusion

We use the GAXRD for the analysis of microstructures of $BaF_2$ thin films grown on Glass, Si and Al substrates. From the GAXRD results shown in Fig. 1. we observe show that the films are showing reflections at 2θ values of $24^0$, $28^0$, $41^0$ and $49^0$ corresponding to (111), (200), (220) and (311) planes of $BaF_2$ respectively. By the analyses of GAXRDresults, one can observe that the deposited films have polycrystalline nature.The crystallite sizes (d) of the samples have been calculated using the Debye Scherrer's formula [18] given by $d = k\lambda / \beta \cos\theta$, where k is shape factor, $\lambda$ is the wavelength of the X-ray source, $\beta$ is full width at half maxima (FWHM) of the peak, θ is the angle corresponding to that peak. We observed that crystallite sizes of glass, Si, and Al are 15.5 nm, 12.4 nm and 15.3 nm respectively.The $R_a$ and $w$ or RMS roughness is computed using the method illustrated in section 2. The computed values of $R_a$ and $w$ are listed in Table 1.

Table 1: The parametric values of average roughness ($R_a$), interface width ($w$), crystallite sizes (d), fractal dimension ($D_f$) and Hurst exponent (H) are listed below

| Substrate | $R_a$ nm | $w$ Nm | d nm | Row | | Column | |
|---|---|---|---|---|---|---|---|
| | | | | $D_f$ | H | $D_f$ | H |
| Glass | 3.39 | 4.02 | 15.5 | 1.32 ± 0.03 | 0.68 ± 0.03 | 1.31 ± 0.04 | 0.69 ± 0.04 |
| Si | 2.41 | 3.02 | 12.4 | 1.61 ± 0.04 | 0.39 ± 0.04 | 1.67 ± 0.04 | 0.33 ± 0.04 |
| Al | 4.86 | 5.68 | 15.3 | 1.40 ± 0.04 | 0.61 ± 0.04 | 1.31 ± 0.04 | 0.69 ± 0.04 |

Fig. 4 shows the surface morphology characterization done with the help of AFM images of $BaF_2$ thin films deposited on all the threesubstrates having thickness 20nm each by electron beam evaporation technique under vacuum environment at room temperature. The roughness parameters ($R_a$ and $w$) give the scale dependent behavior.In order tostudy the different microstructures of the $BaF_2$ thin films surfaces on all the three substrates, we employed the fractal dimensiontechnique to investigate the physical characteristic of AFM images. The fractal dimension is computed for each AFM image in the fast scan direction of 512 rows(columns) traces.For this purpose, the graphs are plotted between the average values of curve length *L(k)* versus *k* at double logarithmic scale as shown inFig.3. The slope obtained by least-squares linear best-fitting method givesthe value of fractal dimension ($D_f$). The calculated values of each film interms of rows and columns are shown in Fig.5. The average value of the fractal dimension is listed in Table 1.

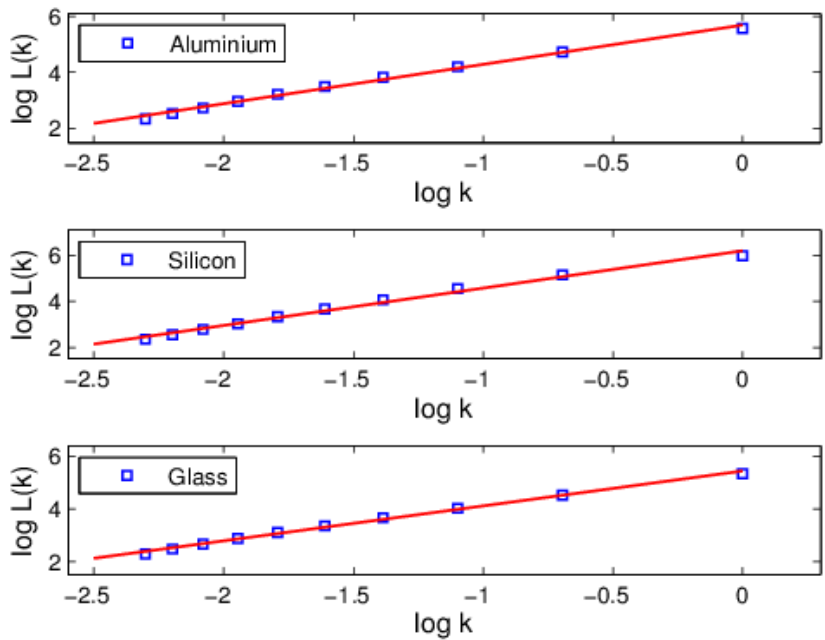

Figure 3: log*L (k)* as function of log *k*. The solid red lines are best-fit curve.

The higher value of fractal dimension suggests that the morphology is more jagged, while the lower value of the fractal dimension corresponds to a smooth surface [5].

We have calculated the fractal dimension of different substrates in terms of function of row and column pixels.Wefound that the fractal dimensions arevery close.So, we can say that the surface shows the isotropic growth[11].The value of Hurst exponent is computed by using the relation $H = 2 - D_f$ [9].The computed values of $H$ are listed in Table 1.It is observed that the value of Hurst exponent is varying between ranges $0 < H < 1$ [12-15]. Here,the values of H for Si and glass aregreater than 0.5 shows that there exists memory effect during deposition[10, 17]whereas for Al the value of Hurst exponent < 0.5 indicates the anti-persistence behavior of the system. The value H = 0.5 indicates that the time series is independent and the system obeys the random walk [10,17].Thus, by analyzing the Hurst exponent and fractal dimension for rows(columns) we can say that for glass substrate the isotropicity nature of BaF$_2$ thin film is highest, whereas for Si it decreases and for Al it is lowest.

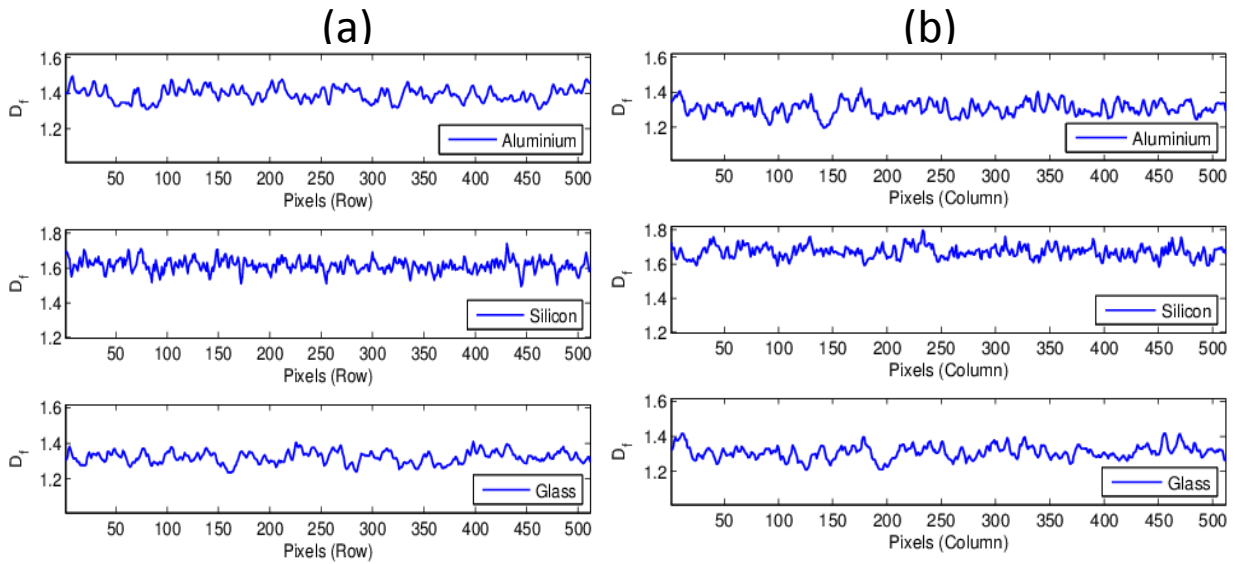

Figure 4: (a) Fractal dimension ($D_f$) value for row pixels for the BaF$_2$ thin film onto the glass, Si and Al substrates.(b) Fractal dimension ($D_f$) value for column pixels for the BaF$_2$ thin film onto the glass, Si and Al substrates.

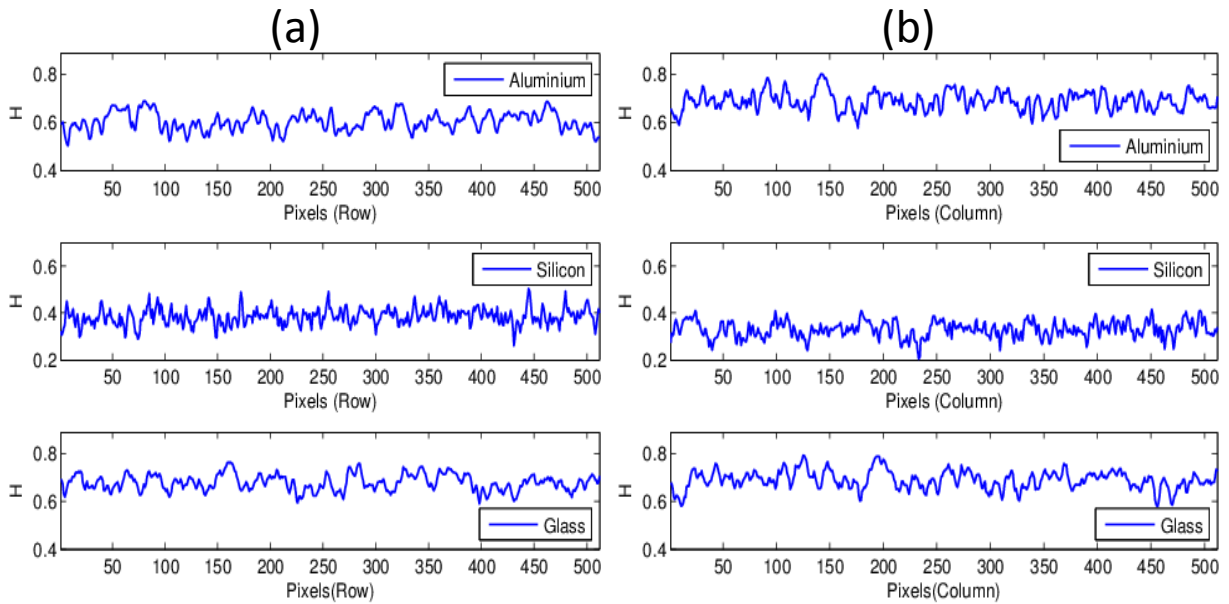

Figure 5: (a) Hurst exponent (*H*) value for row pixels for the BaF$_2$ thin film onto the glass, Si and Al substrates. (b) Hurst exponent (*H*) value for column pixels for the BaF$_2$ thin film onto the glass, Si and Al substrates.

## 4. Conclusion

We studied the surface morphologies of $BaF_2$ thin films deposited by the electron beam evaporation technique onglass, Si and Al substrates. The surface morphology was characterized by using the AFM. The average roughness and interface width have been calculated for each thin film of $BaF_2$ with different substrate. Higuchi's algorithm technique was used to determine the fractal properties of the thin film surface. It was found that the surface roughness and Hurst exponent values are different for different substrates.So, we can conclude that the substrates have great effect on the complexity of the surface. The values of fractal dimensions and Hurst exponents with different orientations revel that the surfaces are isotropic in nature. Further, by finding the values of Hurst exponents and fractal dimensions it is clear that the $BaF_2$ thin film on glass is showing highly isotropic nature than Si and Al.So, we can conclude that the highly isotropic substrate give rise to highly isotropic film.

## Acknowledgements

RPY is thankful to the Science and Engineering Research Board (SERB) of India for awarding the NationalPostdoctoral Fellowship (PDF/2015/000590). We are thankful to Dr. H. P. Bhaskerand Ms. Shama Praveen for their fruitful discussions andvaluable suggestions.

# References

[1] Yadav RP, Pandey RK, Mittal AK, Pandey AC, Chaos. 25 (2015) 083115.

[2] Gibbs W E K, Butterfield AW, Appl. Opt. 14 (1975) 3043-3046.

[3] Țalu S, Bramowicz M, Kulesza S, Ghaderi A, Dalouji V, Solaymani S, FathiKenari M and Ghoranneviss M, Journal of Microscopy 264 (2016)143-152.

[4] Țalu S, Bramowicz M, Kulesza S, Dalouji V, Solaymani S, Valedbagi S, Microscopy Research and Technique 79 (2016) 1208-1213.

[5] Yadav RP, Kumar M, Mittal AK, Dwivedi S, Pandey AC, Materials Let. 126 (2014) 123-125.

[6] Tatom, Frank B, Fractals, 03 (1995) 217-229.

[7] Higuchi T, Physica D, 31 (1988)277–283.

[8] Yadav R P, Agarwal D C, Kumar M, Rajput P, Tomar D S, Pandey S N, Priya PK, Mittal A K, Appl. Surf. Sci. 416 (2017) 51–58.

[9] Yadav RP, Pandey RK, Mittal AK, Kumar M, Pandey A C, Materials Focus, 4 (2015) 403-408.

[10] Li J M, Image-Based Fractal Description Of microstructures, Springer Science Business Media New York; 2003.

[11] Alvarez-R J, Echeverria J C, Rodriguez E, Physica A 387 (2008) 6452–6462.

[12] Feder J, Fractals, Plenum Press, New York, 1988.

[13] Saitou M, Oshikawa W, Makabe A, J of Phy and Chem of Solids 63 (2002)1685–1689.

[14] Ferreira S O, Ribeiro I R B, Suela J, Menezes S I L, Ferreira SC, Appl. Phys. Lett. 88 (2006) 244102.

[15] Mandelbrot B B. PhysicaScripta 32 (1985) 257-260.

[16] Yadav R P, Kumar T, Baranwal V, Vandana, M Kumar, Priya P K, Pandey S N, Mittal A K, J. Appl. Phys. 121 (2017) 0553011-0553017.

[17] Mandelbrot B B, The Fractal Geometry of Nature, Freemen Press, San Francisco (1982).

[18] Cullity BD, Elements of X-ray diffraction, Addition-Wesley Publishing Company, Inc.; 1956.